\documentclass[3p,times,procedia]{elsarticle}
\usepackage{nupha_ecrc}

\volume{00}

\firstpage{1}

\journalname{Nuclear Physics A}

\runauth{}

\jid{nupha}

\jnltitlelogo{Nuclear Physics A}

\usepackage{amssymb}

\usepackage[figuresright]{rotating}
\usepackage{amssymb}   
\usepackage{amsmath}
\usepackage[braket, qm]{qcircuit}
\usepackage{hyperref}

\newcommand{\Eq}[1]{Eq.~(\ref{#1})}
\newcommand{\Eqs}[2]{Eqs.~(\ref{#1}-\ref{#2})}

\begin{document}

\begin{frontmatter}



\dochead{XXVIIIth International Conference on Ultrarelativistic Nucleus-Nucleus Collisions\\ (Quark Matter 2019)}

\title{Computing real time correlation functions\\on a hybrid classical/quantum computer}


\author[NM]{Niklas Mueller}
\author[NM,AT]{Andrey Tarasov}
\author[NM]{Raju Venugopalan}

\address[NM]{Physics Department, Brookhaven National Laboratory, Bldg. 510A, Upton, NY 11973, USA}
\address[AT]{Department of Physics, The Ohio State University, Columbus, OH 43210-1117, USA}

\begin{abstract}
Quantum devices may overcome limitations of classical computers in studies of nuclear structure functions and parton Wigner distributions of protons and nuclei.
In this talk, we discuss a worldline approach to compute nuclear structure functions in the high energy Regge limit of QCD using a hybrid quantum computer, by expressing the fermion determinant in the QCD path integral as a quantum mechanical path integral over $0+1$-dimensional fermionic and bosonic world-lines in background gauge fields. Our simplest example of computing the well-known dipole model result for the structure function $F_2$ in the high energy Regge limit is feasible with NISQ era technology using few qubits and shallow circuits. This example can be
scaled up in complexity and extended in scope to compute structure functions, scattering amplitudes and other real-time correlation functions in QCD, relevant  for example to describe non-equilibrium transport of quarks and gluons in a Quark-Gluon-Plasma.
\end{abstract}


\end{frontmatter}


\section{Introduction}\label{sec:Introduction}
Classical first principles computations of nuclear structure functions and parton Wigner distributions is an outstanding problem in Quantum Chromodynamics (QCD), and an attractive candidate to explore a potential 
computational opportunity for present and future quantum devices.

QCD structure functions involve non-perturbative nucleon/nuclear matrix elements of electromagnetic currents that are light-like separated in Minkowskian spacetime and thus difficult to compute 
with classical lattice Monte Carlo techniques which are restricted to Euclidean spacetime. Important developments are the Operator Product Expansion (OPE) \cite{Winter:2017bfs} to compute moments of structure functions on the lattice,
and  quasi- or pseudo parton distribution functions (pdf's) \cite{Ji:2013dva,Radyushkin:2016hsy}.

Quantum computers and analog simulators may also overcome the limitations of classical algorithms in other aspects relevant for ultra-relativistic heavy ion collisions. Examples are the thermodynamic properties of the Quark-Gluon-Plasma (QGP)
at finite chemical potential in QCD \cite{Ortiz:2000gc,Alexandru:2019ozf}, and real-time correlation functions for transport phenomena at weak and strong coupling. At present, Noisy-Intermediate-Scale Quantum (NISQ) era digital quantum computers and analog quantum simulators offer limited resources and error-tolerance,  requiring efficient digitization strategies, when mapping quantum field theories onto quantum mechanical systems~\cite{cirac2012goals,hauke2012can,nuclearReview}.  

In this talk \cite{Mueller:2019qqj}, we will present a novel worldline approach \cite{Strassler:1992zr,DHoker:1995uyv,Mueller:2017arw,Mueller:2017lzw} to computing proton and nuclear structure functions measured in deeply inelastic scattering (DIS) of electrons off nuclear targets~\cite{Breidenbach:1969kd,Bjorken:1968dy}, whereby the fermion determinant of the QCD Schwinger-Keldysh effective action is expressed as a quantum mechanical path integral of $0+1$ dimensional bosonic and fermionic worldlines.
In this formulation, the problem can be expressed as a von Neumann problem of a combined quark worldline and Yang-Mills density matrix. Aiming at problems that are feasible on present and near-future devices, we restrict
ourselves to the computation of the structure function $F_2$ in the high energy `Regge limit', where the computation is significantly simpler. We will present here a hybrid approach where only part of the computation is done on a quantum
computer with few qubits and shallow circuit depth.

\section{Worldline Approach}\label{sec:worldlineapproach}
In the worldline formalism, one expresses the Euclidean QCD+QED fermion effective action as a quantum mechanical path integral of bosonic position and momentum variables $x_\mu(\tau)$, $p_\mu(\tau)$ ($\mu=0,\dots,3$), as well
as Grassmann variables $\theta_i(\tau), \theta_i^*(\theta)$ ($i=1,2$), to represent the internal Dirac structure (a similar Grassmann representation also exists for color \cite{Mueller:2019gjj}). Analytically continuing this to Minkowskian space time, 
one obtains the following effective action \cite{Mueller:2019qqj,Tarasov:2019rfp},
\begin{align}\label{eq:effectiveactionWL}
\Gamma[A,a]=-\frac{i}{2}\int\limits_0^\infty \frac{dT}{T} \text{tr}_c \int_{\rm P} \mathcal{D}x\mathcal{D}p \int_{\rm AP} \mathcal{D}\theta \mathcal{D}\theta ^* e^{iS}\,,\qquad S\equiv \int_0^T d\tau ( p_\mu \dot{x}^\mu -\frac{i}{2}\dot{\theta}_i \theta_i^* + \frac{i}{2}\theta_i \dot{\theta}^*_i - H)\,,
\end{align}
where the Hamiltonian is $H\equiv P^2 + ig \psi^\mu F_{\mu\nu}[A]\psi^\nu + ie\psi^\mu F_{\mu\nu}[a]\psi^\nu$ and $\psi^0=(\theta_1^*-\theta_1 )/\sqrt{2}$, $\psi^3=(\theta_1^*+\theta_1)/\sqrt{2}$, $\psi^1=(\theta_2^*+\theta_2)/\sqrt{2}$ and
$\psi^3=-i(\theta_1^*+\theta_1)/\sqrt{2}$, ${\rm P}$ (${\rm AP}$) denote periodic (antiperiodic) boundary conditions on the closed worldline of length $T$ for bosons (fermions).


\section{DIS in the Regge limit}\label{sec:DIS}
In the inclusive scattering process of an electron ($\ell_e$) off a proton $N$, $\ell_e(l)+N(P)\rightarrow \ell_e(l')+X$, the cross section can be factorized  into a perturbatively computable lepton- and a non-perturbative
hadron tensor, 
\begin{align}\label{eq:Wdefinition}
W_{\mu\nu}(q,P,S)=\text{Im} \frac{i}{\pi}\int d^4\mathbf{x}\, e^{i\mathbf{q}\cdot \mathbf{x}} \,\langle P,S|\, \mathbb{T} \hat{j}^\mu(\mathbf{x}) \hat{j}^\nu (0)\, | P,S\rangle\,,
 \end{align}
expressed by a non-perturbative matrix element (of the proton density matrix) of ($\mathbb{T}$-ordered) electric current operators $\hat{j}^\mu$. Here $q$ is the four momentum of the exchanged virtual photon and $P,S$ are the proton's momentum and spin, respectively. 
In \cite{Mueller:2019qqj}, we used a worldline representation of the QCD Schwinger-Keldysh path integral to write \Eq{eq:Wdefinition} as
\begin{align}\label{eq:masterformula}
W^{\mu\nu} = & \frac{1}{\pi e^2}\text{Im}\,  \int d^4 z \, e^{iq\cdot z}\sum\limits_{n=0}^\infty  \frac{i^{n+4}}{n!} \int  \Big[\prod\limits_{k=1}^{n+4} d^4x_1^k d^4x_2^k 
d^2\theta_1^k d^2\theta_2^k  \Big]
  \int dA_1  dA_2\;  \,  \text{tr}_c\;
\langle  x_1, -\theta_1, A_1 | \hat{\rho}_\text{init} | x_2 ,\theta_2, A_2 \rangle
\nonumber\\
&\times  \langle x_2,\theta_2, A_2 |   \hat{\mathbb{U}}_{(-\infty,z)} \hat{J}^\mu_{(4)}(z)  \hat{\mathbb{U}}_{(z,\infty)}  
\hat{\mathbb{U}}_{(\infty,0)}  \hat{J}_{(4)}^\nu(0) \hat{\mathbb{U}}_{(0,-\infty)} | x_1,\theta_1, A_1\rangle\,,
\end{align}
where $|x,\theta,A\rangle=|x,\theta\rangle |A\rangle$, $|x,\theta\rangle = \prod_{k=1}^3 |x^k,\theta^k\rangle$ and  $\hat{\mathbb{U}}_{(t,t')}  \equiv \exp{\{-i  \hat{H}(t-t')\}}$ is the worldline and Yang-Mills evolution
operator with $\hat{H}= \hat{H}_\text{YM}  +  \sum_{k=1}^{4+n}  \hat{H}^k $ the sum of $k=1,\dots,N$ worldline Hamiltonians given in \Eq{eq:effectiveactionWL} (in equal-time quantization)  and the Yang-Mills Hamiltonian in temporal-axial gauge.
Here, $\hat{J}_{(4)}^\mu(z)\equiv \sum{}_{k=1}^4 \hat{j}_k^\mu(z)$ is the electromagnetic current operator of the three valence ($k=1,2,3$) and sea-quarks $k=4$ with
\begin{align}
\hat{j}_k^\mu(z)\equiv \frac{e}{p_k^0} \Big[\hat{P}_k^\mu + i \hat{\psi}_k^\nu\hat{\psi}_k^\mu q_\nu \Big] \delta^{(3)}(\mathbf{z}-\hat{\mathbf{x}}_k(z^0) )\,,
\end{align}
obtained by varying the worldline Hamiltonian by an electromagnetic field $a^\mu$(z). Further, $ \hat{\rho}_\text{init}$ is the proton's density matrix, carrying the proton's quantum numbers, prepared in the past in terms of three valence worldline quarks
and with the interactions switched off. To adiabatically prepare the (fully interacting) proton state, one solves a combined Yang-Mills and worldline Hamiltonian operator equation $\partial_t \hat{\rho}=-i[\hat{H},\hat{\rho}]$ with the initial condition $ \hat{\rho}_\text{init}$ on a quantum computer.

Evolving a combined worldline/Yang-Mills proton density matrix and solving \Eq{eq:masterformula} is difficult with presently available NISQ era quantum devices. This is particularly true for the Yang-Mills part, where
efforts are focussed on realizations of the Kogut-Susskind lattice Hamiltonian \cite{Kogut:1974ag}, see e.g. \cite{Banerjee:2012xg,Anishetty:2009nh,Zohar:2012xf}.

 To make progress, we resort to a much simpler approach, which is plausbile in the high-energy ``Regge" limit. This corresponds to a fixed photon virtuality $Q^2=-q^2$, and small Bjorken $x_{\rm Bj}\approx Q^2/s\rightarrow0$, where $s\approx 2P^+q^-\rightarrow \infty$ is the squared center of mass energy with $P^+=(P^0+P^3)/\sqrt{2}$ ($q^-=(q^0-q^3)/\sqrt{2}$) the light-cone momentum of a right-moving proton (left-moving photon). The slow $x_{\rm Bj}\ll 1$ degrees of freedom can be approximated by dynamical classical color gauge fields coupled to static color sources at $x_{\rm Bj}\sim 1 $, as formalized in the Color Glass Condensate effective field theory (CGC EFT) \cite{McLerran:1993ni,Gelis:2010nm}. In  this EFT,  the typical solutions are classical `shockwave' color fields, sharply localized in the lightcone coordinate $x^-=(x^0-x^3)/\sqrt{2}$.  We study
 the projection of the hadron tensor $F_2 \equiv \Pi_2^{\mu\nu} W_{\mu\nu}$, where $\Pi_2^{\mu\nu}\equiv \frac{3P\cdot q}{4a}[\frac{P^\mu P^\nu}{a}-\frac{g^{\mu\nu}}{3}]$, $a=P\cdot q/(2 x_{\rm Bj})+M^2$, and $M$ is the hadron mass.  In the CGC, 
it can be written as
\begin{align}\label{eq:F2}
&F_2(q,P)=
\frac{\sigma \, Q^2 }{2\pi e^2} \int [\mathcal{D} \rho] W[\rho]  \int\limits_{x_\perp} \int\limits_z  \sum\limits_{L,T;\, f} |\Psi^f_{L,T}(z, x_\perp)|^2
 D_\rho(x_\perp) \, i \int d^2\theta \langle -\theta | \big[ {\Omega}_{L,T}(z, x_\perp)\big] | \theta \rangle\,,
\end{align}
where $\mathcal{D}\rho$ is a functional integral over large $x_{\rm Bj}$ sources, $D_\rho(x_\perp) $ is the dipole amplitude for a given $\rho$, $|\Psi_{L/T}|^2(x_\perp,z)$ is the virtual $\gamma^*\rightarrow q\bar{q}$ squared wavefunction with transverse size $x_\perp$ and quark lightcone momentum fraction $z\equiv p^-/q^-$. These are analytically known or computed classically (see \cite{Mueller:2019qqj}). 
The Grassmann integral in \Eq{eq:F2} represents the operator trace with worldline operator
\begin{align}\label{eq:omega}
&\Omega_L(z,x_\perp)= \frac{1}{ 2z(1-z) } \Big\{ - \frac{3}{4} ~ [ ( 2z - 1 ) + 2 \hat{\psi}^- \hat{\psi}^+ ] 
[ (2z - 1 ) - 2 \hat{\psi}^- \hat{\psi}^+ ] - \hat{\psi}^+ \hat{\psi}^-   \hat{\psi}^+ \hat{\psi}^-  -   \hat{\psi}^j \hat{\psi}^+ \hat{\psi}^j \hat{\psi}^- 
- z(1 - z) + \frac{3}{4} \Big\}\,,
\end{align}
where $\psi^\pm \equiv (\psi^0\pm \psi^3)/\sqrt{2}$, and $\Omega_T(z,x_\perp)=1$. 

\section{Quantum Circuits}
\Eqs{eq:F2}{eq:omega} can be evaluated on a quantum computer, quantizing $\psi^\mu \rightarrow \hat{\psi}^\mu= \gamma_5\gamma^\mu/\sqrt{2}$ where $\gamma^\mu $ are the Dirac matrices in Minkowskian metric, $[\gamma_\mu,\gamma_\nu]_+=2g_{\mu\nu}$ ($g=\text{diag}(+,-,-,-)$) and $\gamma_5=i\gamma^0\gamma^1\gamma^2\gamma^3$. On can express these in terms of fermion creation and annihilation operators $\hat{b}_i^\dagger, \hat{b}_i$ ($i=1,2$), using
$\hat{\psi}^0 = {(\hat{b}_1^\dagger -\hat{b}_1 })/{\sqrt{2}}$, $\hat{\psi}^3 = {(\hat{b}_1^\dagger +\hat{b}_1) }/{\sqrt{2}}$, $\hat{\psi}^1 = (\hat{b}_2^\dagger +\hat{b}_2 )/\sqrt{2}$, and  $\hat{\psi}^2 = -i(\hat{b}_2^\dagger -\hat{b}_2 )/\sqrt{2}$,
which satisfy the usual anticommutation relations $[\hat{b}_i^\dagger,\hat{b}_j]_+ = \delta_{ij}$. By means of a Jordan-Wigner transformation 
$\hat{b}_1^\dagger = (\sigma^x- i\sigma^y)/2\otimes \mathbb{I}$, $\hat{b}_1 = (\sigma^x+ i\sigma^y)/2\otimes \mathbb{I}$,  $\hat{b}_2^\dagger = \sigma^z\otimes (\sigma^x- i\sigma^y)/2$ and $\hat{b}_2 = \sigma^z\otimes (\sigma^x+ i\sigma^y)/2$
 one can then write the terms in \Eq{eq:omega} as
\begin{align}\label{eq:photonvertex}
\hat{\psi}^- \hat{\psi}^+ = -\frac{1}{2}\,  [\mathbb{I}+ \sigma^z] \otimes \mathbb{I}\,,\qquad
\hat{\psi}^1 \hat{\psi}^\pm = - \frac{1}{2\sqrt{2}} \, [\sigma^x\mp i \sigma^y] \otimes \sigma^x \,,\qquad
\hat{\psi}^2 \hat{\psi}^\pm &=  \frac{1}{2\sqrt{2}}\,  [\sigma^x\mp i \sigma^y] \otimes \sigma^y \,,
\end{align}
where $\sigma^i$ are standard gates. The spin trace in \Eq{eq:F2} can now be written as the following ($n=2$) quantum circuit~\cite{knill1998power,datta2005entanglement}
\begin{align} \label{eq;tracecircuit}
\Qcircuit @C=.5em @R=-.5em @!R {
&\lstick{ \hat{\rho}_c = | 0 \rangle \langle 0 |}   & \gate{H} &  \ctrl{1} & \qw &  \qw & \meterB{ \sigma }\inputgroup{2}{3}{.95em}{\hat{\rho}_n= \mathbb{I}_n /2^n } \\
& &  \lstick{  }& \multigate{1}{{\Omega}_{L,T}} & \qw  & \qw \\
 & & \lstick{ (n \text{ qubits})} & \ghost{{\Omega}_{L,R}} & \qw & \qw
}\,,
\end{align}
involving the controlled gate of $\Omega$, further $\hat{\rho}_c$ is the density matrix of an auxiliary qubit and $\hat{\rho}_n$ is a maximally mixed state of $n$ qubits, and $H$ is the Hadamard gate. Measurements of the Pauli operators $\langle \sigma^x \rangle$ and $\langle \sigma^y \rangle$ yield the real and imaginary part of $\text{Tr}[\Omega_{L/T}]$, respectively \cite{datta2005entanglement}.

\section{Conclusions}\label{sec:Conclusions}
We presented a worldline approach to explore a potential quantum opportunity in computing structure functions in QCD on present and future quantum computers. Our hybrid algorithm is part of a bottom-up approach towards ultimately computing the structure of protons and nuclei from first principles. The hybrid worldline formulation is a simple enough starting point to be implemented on currently available, noisy and small quantum devices.
Extending the program laid out in \cite{Jordan:2011ne} to gauge theories, one can hope to compute not only structure functions but also multi-leg and multi-loop scattering amplitudes~\cite{Bern:1991aq,Bern:1994cg} using this Hamiltonian formulation. One particular benefit is a well defined prescription for regulating UV divergences. 

In heavy ion collisions, quantum computation may be useful to study non-equilibrium transport phenomena; an example that closely resembles our discussion is photon production in the quark-gluon plasma~\cite{McLerran:1984ay}.

\textit{Acknowledgements.} The authors are supported by the U.S. Department of Energy, Office of Science, Office of Nuclear Physics, under contract No. DE- SC0012704, within the framework of the Beam Energy Scan Theory (BEST) Topical Collaboration and the Topical Collaboration for the Coordinated Theoretical Approach to Transverse Momentum Dependent Hadron Structure in QCD (TMD Collaboration).  NM is funded by the Deutsche Forschungsgemeinschaft (DFG, German Research Foundation) - Project 404640738.  AT is supported by U.S. DOE grant DE-SC0004286 and Center for Frontiers in Nuclear Science (CFNS) at Stony Brook University and Brookhaven National Laboratory.





\bibliographystyle{elsarticle-num}
\bibliography{references}

\begin{thebibliography}{10}
\expandafter\ifx\csname url\endcsname\relax
  \def\url#1{\texttt{#1}}\fi
\expandafter\ifx\csname urlprefix\endcsname\relax\def\urlprefix{URL }\fi
\expandafter\ifx\csname href\endcsname\relax
  \def\href#1#2{#2} \def\path#1{#1}\fi

\bibitem{Winter:2017bfs}
F.~Winter~\textit{et al.}, Phys. Rev. D96~(9) (2017) 094512.

\bibitem{Ji:2013dva}
X.~Ji, Phys. Rev. Lett. 110 (2013) 262002.
\newblock \href {http://dx.doi.org/10.1103/PhysRevLett.110.262002}
  {\path{doi:10.1103/PhysRevLett.110.262002}}.

\bibitem{Radyushkin:2016hsy}
A.~Radyushkin, Phys. Lett. B767 (2017) 314--320.
\newblock \href {http://dx.doi.org/10.1016/j.physletb.2017.02.019}
  {\path{doi:10.1016/j.physletb.2017.02.019}}.

\bibitem{Ortiz:2000gc}
G.~Ortiz, J.~E. Gubernatis, E.~Knill, R.~Laflamme, Phys. Rev. A64 (2001)
  022319, [Erratum: Phys. Rev.A65,029902(2002)].

\bibitem{Alexandru:2019ozf}
A.~Alexandru, P.~F. Bedaque, H.~Lamm, S.~Lawrence, Phys. Rev. Lett. 123~(9)
  (2019) 090501.

\bibitem{cirac2012goals}
J.~I. Cirac, P.~Zoller, Nature Physics 8~(4) (2012) 264.

\bibitem{hauke2012can}
P.~Hauke, F.~M. Cucchietti, L.~Tagliacozzo, I.~Deutsch, M.~Lewenstein, Reports
  on Progress in Physics 75~(8) (2012) 082401.

\bibitem{nuclearReview}
J.~Carlson, D.~Dean, M.~Hjorth-Jensen, D.~Kaplan, J.~Preskill, K.~Roche,
  M.~Savage, M.~Troyer, INT Report 18-008.

\bibitem{Mueller:2019qqj}
N.~Mueller, A.~Tarasov, R.~Venugopalan, {}\href
  {http://arxiv.org/abs/1908.07051} {\path{arXiv:1908.07051}}.

\bibitem{Strassler:1992zr}
M.~J. Strassler, Nucl. Phys. B385 (1992) 145--184.
\newblock \href {http://arxiv.org/abs/hep-ph/9205205}
  {\path{arXiv:hep-ph/9205205}}, \href
  {http://dx.doi.org/10.1016/0550-3213(92)90098-V}
  {\path{doi:10.1016/0550-3213(92)90098-V}}.

\bibitem{DHoker:1995uyv}
E.~D'Hoker, D.~G. Gagne, Nucl. Phys. B467 (1996) 297--312.
\newblock \href {http://dx.doi.org/10.1016/0550-3213(96)00126-5}
  {\path{doi:10.1016/0550-3213(96)00126-5}}.

\bibitem{Mueller:2017arw}
N.~Mueller, R.~Venugopalan, {}, Phys. Rev. D96~(1) (2017) 016023.
\newblock \href {http://arxiv.org/abs/1702.01233} {\path{arXiv:1702.01233}},
  \href {http://dx.doi.org/10.1103/PhysRevD.96.016023}
  {\path{doi:10.1103/PhysRevD.96.016023}}.

\bibitem{Mueller:2017lzw}
N.~Mueller, R.~Venugopalan, {}, Phys. Rev. D97~(5) (2018) 051901.
\newblock \href {http://arxiv.org/abs/1701.03331} {\path{arXiv:1701.03331}},
  \href {http://dx.doi.org/10.1103/PhysRevD.97.051901}
  {\path{doi:10.1103/PhysRevD.97.051901}}.

\bibitem{Breidenbach:1969kd}
M.~Breidenbach, J.~I. Friedman, H.~W. Kendall, E.~D. Bloom, D.~H. Coward, H.~C.
  DeStaebler, J.~Drees, L.~W. Mo, R.~E. Taylor, Phys. Rev. Lett. 23 (1969)
  935--939.
\newblock \href {http://dx.doi.org/10.1103/PhysRevLett.23.935}
  {\path{doi:10.1103/PhysRevLett.23.935}}.

\bibitem{Bjorken:1968dy}
J.~D. Bjorken, {}, Phys. Rev. 179 (1969) 1547--1553.
\newblock \href {http://dx.doi.org/10.1103/PhysRev.179.1547}
  {\path{doi:10.1103/PhysRev.179.1547}}.

\bibitem{Mueller:2019gjj}
N.~Mueller, R.~Venugopalan, Phys. Rev. D99~(5) (2019) 056003.
\newblock \href {http://arxiv.org/abs/1901.10492} {\path{arXiv:1901.10492}},
  \href {http://dx.doi.org/10.1103/PhysRevD.99.056003}
  {\path{doi:10.1103/PhysRevD.99.056003}}.

\bibitem{Tarasov:2019rfp}
A.~Tarasov, R.~Venugopalan, Phys. Rev. D100~(5).
\newblock \href {http://dx.doi.org/10.1103/PhysRevD.100.054007}
  {\path{doi:10.1103/PhysRevD.100.054007}}.

\bibitem{Kogut:1974ag}
J.~B. Kogut, L.~Susskind, Phys. Rev. D11 (1975) 395--408.
\newblock \href {http://dx.doi.org/10.1103/PhysRevD.11.395}
  {\path{doi:10.1103/PhysRevD.11.395}}.

\bibitem{Banerjee:2012xg}
D.~Banerjee, M.~Bögli, M.~Dalmonte, E.~Rico, P.~Stebler, U.~J. Wiese,
  P.~Zoller, {}, Phys. Rev. Lett. 110~(12).

\bibitem{Anishetty:2009nh}
R.~Anishetty, M.~Mathur, I.~Raychowdhury, {}, J. Phys. A43 (2010) 035403.
\newblock \href {http://dx.doi.org/10.1088/1751-8113/43/3/035403}
  {\path{doi:10.1088/1751-8113/43/3/035403}}.

\bibitem{Zohar:2012xf}
E.~Zohar, J.~I. Cirac, B.~Reznik, Phys. Rev. Lett. 110~(12) (2013) 125304.
\newblock \href {http://dx.doi.org/10.1103/PhysRevLett.110.125304}
  {\path{doi:10.1103/PhysRevLett.110.125304}}.

\bibitem{McLerran:1993ni}
L.~D. McLerran, R.~Venugopalan, Phys. Rev. D49 (1994) 2233--2241.
\newblock \href {http://dx.doi.org/10.1103/PhysRevD.49.2233}
  {\path{doi:10.1103/PhysRevD.49.2233}}.

\bibitem{Gelis:2010nm}
F.~Gelis, E.~Iancu, J.~Jalilian-Marian, R.~Venugopalan, Ann. Rev. Nucl. Part.
  Sci. 60 (2010) 463--489.

\bibitem{knill1998power}
E.~Knill, R.~Laflamme, Physical Review Letters 81~(25) (1998) 5672.

\bibitem{datta2005entanglement}
A.~Datta, S.~T. Flammia, C.~M. Caves, Physical Review A 72~(4) (2005) 042316.

\bibitem{Jordan:2011ne}
S.~P. Jordan, K.~S.~M. Lee, J.~Preskill, Science 336 (2012) 1130--1133.
\newblock \href {http://dx.doi.org/10.1126/science.1217069}
  {\path{doi:10.1126/science.1217069}}.

\bibitem{Bern:1991aq}
Z.~Bern, D.~A. Kosower, Nucl. Phys. B379 (1992) 451--561.
\newblock \href {http://dx.doi.org/10.1016/0550-3213(92)90134-W}
  {\path{doi:10.1016/0550-3213(92)90134-W}}.

\bibitem{Bern:1994cg}
Z.~Bern, L.~J. Dixon, D.~C. Dunbar, D.~A. Kosower, Nucl. Phys. B435 (1995)
  59--101.
\newblock \href {http://dx.doi.org/10.1016/0550-3213(94)00488-Z}
  {\path{doi:10.1016/0550-3213(94)00488-Z}}.

\bibitem{McLerran:1984ay}
L.~D. McLerran, T.~Toimela, Phys. Rev. D31 (1985) 545.
\newblock \href {http://dx.doi.org/10.1103/PhysRevD.31.545}
  {\path{doi:10.1103/PhysRevD.31.545}}.

\end{thebibliography}







\end{document}